\documentclass[12pt]{article}
\usepackage{amssymb}
\usepackage{epsfig}

\catcode`\@=11
\def\numberbysection{\@addtoreset{equation}{section}
        \def\theequation{\thesection.\arabic{equation}}}
\def\half{\frac{1}{2}}
\def\beq{\begin{equation}}
\def\eeq{\end{equation}}
\numberbysection
\begin{document}
\begin{titlepage}
\begin{center}
\hfill DFF  1/10/02 \\
\vskip 1.in {\Large \bf Projective modules over the fuzzy
four-sphere} \vskip 0.5in P. Valtancoli
\\[.2in]
{\em Dipartimento di Fisica, Polo Scientifico Universit\'a di Firenze \\
and INFN, Sezione di Firenze (Italy)\\
Via G. Sansone 1, 50019 Sesto Fiorentino, Italy}
\end{center}
\vskip .5in
\begin{abstract}
We describe how to reduce the fuzzy four-sphere algebra to a set
of four independent raising and lowering oscillator operators. In
terms of them we derive the projector valued operators for the
fuzzy four-sphere, which are the global definition of
$k$-instanton connections over this noncommutative base manifold.
\end{abstract}
\medskip
\end{titlepage}
\pagenumbering{arabic}
\section{Introduction}
In ref. \cite{12} ( see also \cite{7} ) a finite projective module
description of all noncommutative monopole configurations on the
fuzzy two-sphere has been presented. The basic variables which
describe connections and gauge fields of the monopoles are
represented by globally defined projective operators, in terms of
which it is possible to deduce directly the Chern class
\cite{1}-\cite{2}-\cite{3}-\cite{4}.

The aim of the present paper is to generalize the above
construction to the case of the fuzzy four-sphere
\cite{8}-\cite{9}-\cite{10}-\cite{11}. The difference from the
fuzzy two-sphere case \cite{5}-\cite{6} is that the commutators of
the coordinates do not close between them in this case. We have to
generalize the algebra from five hermitian operators to fifteen
operators, which define the $SO(5,1)$ algebra. Despite this
difficulty we are able to generalize our method. The starting
point is the simple observation that our construction, in the
classical limit, is based on the Hopf fibration $S^3 \rightarrow
S^2$; since it exists its generalization to the case $S^7
\rightarrow S^4$ we conclude that the fuzzy four-sphere algebra is
generated by four independent oscillators. Once that the building
blocks of the more complicated algebra are recognized, it is then
easy to compute the projectors, which define the projective
modules over the fuzzy four-sphere. The projectors are $N \times
N$ matrices, having as entries the elements of the basic
noncommutative coordinate algebra of the fuzzy four-sphere, where
$N = \frac{(k+1)(k+2)(k+3)}{6}$ and $k$ is an integer, labelling
the instanton number.

It is important to notice that a further generalization to higher
dimensional fuzzy spheres of the present method is forbidden since
there is no straightforward generalization of the Hopf fibration
to $S^{2n}$ spheres. At the level of the fuzzy four-sphere, there
are sixteen independent entries of the type $z_i\overline{z}_j$ (
where ($z_i, \overline{z}_j$) are the basic raising and lowering
operators of the four oscillators ), corresponding to the basic
operators of the $SO(5,1)$ algebra and the number operator
$\hat{N}$, labelling the representations of this algebra.

\section{The fuzzy four-sphere}

The fuzzy four-sphere  ( we follow mainly ref. \cite{8} ) is
realized through a set of five hermitian operators that satisfy
the following two conditions:

\begin{eqnarray}& \ &
\epsilon^{\mu\nu\lambda\rho\sigma} \hat{x}_{\mu} \hat{x}_{\nu}
\hat{x}_{\lambda} \hat{x}_{\rho} = C \hat{x}_{\sigma} \nonumber \\
& \ & \hat{x}_{\mu} \hat{x}_{\mu} = R^2 \label{21}
\end{eqnarray}

where $R$ is a radius of the sphere. The two constraints respect
$SO(5)$ invariance. Both  constraints are for example realized by
the five-dimensional gamma matrices, which are the simplest
finite-dimensional realizations of the fuzzy four-sphere, like the
Pauli matrices are the simplest one for the fuzzy two-sphere.

In general, let us define finite-dimensional matrices
$\hat{\Gamma}_{\mu}$ as follows

\beq \hat{x}_{\mu} = \rho \hat{\Gamma}_{\mu} \label{22} \eeq

where $\rho$ is a dimensional constant. The gamma matrices algebra
is defined as

\beq \hat{\Gamma}^{(0)}_1 = \gamma_0 \ \ \ \ \hat{\Gamma}^{(0)}_2
= \gamma_0 \gamma_1 \ \ \ \ \ \hat{\Gamma}^{(0)}_3 = \gamma_0
\gamma_2 \ \ \ \ \ \hat{\Gamma}^{(0)}_4 = \gamma_0 \gamma_3 \ \ \
\ \hat{\Gamma}^{(0)}_5 = i \gamma_0 \gamma_5 . \label{23} \eeq

The general matrices can be built from the $n$-fold symmetric
tensor product of $\hat{\Gamma}^{(0)}_{\mu}$ algebra :

\beq \hat{\Gamma}^{(n)}_{\mu} = ( \hat{\Gamma}^{(0)}_{\mu} \otimes
1 \otimes .... \otimes 1 + 1 \otimes \hat{\Gamma}^{(0)}_{\mu}
\otimes  .... \otimes 1 + ..... 1 \otimes 1 \otimes .... \otimes
\hat{\Gamma}^{(0)}_{\mu} )_{sym} . \label{24} \eeq

The size of these matrices depends on the parameter $n$ as

\beq N = \frac{(n+1)(n+2)(n+3)}{6} \label{25} .\eeq

The same construction is valid in two dimensions, i.e. replacing $
\hat{\Gamma}^{(0)}_\mu $ with Pauli matrices $
\hat{\Gamma}^{(n)}_\mu $ become the hermitian operators of a fuzzy
two-sphere, that is a $(n+1)$-dimensional representation of
$SU(2)$.

It is not difficult to realize that these matrices satisfy the two
conditions defining the fuzzy four-sphere (\ref{21}) :

\begin{eqnarray} & \ & \hat{\Gamma}^{(n)}_{\mu} \hat{\Gamma}^{(n)}_{\mu} = n ( n +
4 ) \nonumber \\
& \ & \epsilon^{\mu\nu\lambda\rho\sigma} \hat{\Gamma}^{(n)}_{\mu}
\hat{\Gamma}^{(n)}_{\nu} \hat{\Gamma}^{(n)}_{\lambda}
\hat{\Gamma}^{(n)}_{\rho} = \epsilon^{\mu\nu\lambda\rho\sigma}
\hat{\Gamma}^{(n)}_{\mu\nu} \hat{\Gamma}^{(n)}_{\lambda\rho} = (
8n + 16 ) \hat{\Gamma}^{(n)}_{\sigma} \label{26}
\end{eqnarray}

where $ \hat{\Gamma}^{(n)}_{\mu\nu} = \frac{1}{2} [
\hat{\Gamma}^{(n)}_{\mu}, \hat{\Gamma}^{(n)}_{\nu} ] $.

The two parameters $C$ and $R^2$ can be computed as:

\begin{eqnarray}
C & = & ( 8 n + 16 ) \rho^3 \nonumber \\
R^2 & = & n ( n + 4 ) \rho^2 . \label{27}
\end{eqnarray}

In the large $n$-limit, fixing $R^2$, one recovers the usual
classical four-sphere, since then $C$ goes to zero as
$\frac{1}{n^2}$. From the second constraint we can recover the
following useful relation :

\beq \hat{\Gamma}^{(n)}_{\mu\nu} = - \frac{1}{2(n+2)}
\epsilon^{\mu\nu\lambda\rho\sigma}
\hat{\Gamma}^{(n)}_{\lambda\rho} \hat{\Gamma}^{(n)}_{\sigma} = -
\frac{1}{2(n+2)} \epsilon^{\mu\nu\lambda\rho\sigma}
\hat{\Gamma}^{(n)}_{\lambda} \hat{\Gamma}^{(n)}_{\rho}
\hat{\Gamma}^{(n)}_{\sigma} . \label{28} \eeq

For completeness let us give a whole set of relations, obtained
from the fundamental ones (\ref{26}) and (\ref{28}):

\begin{eqnarray}
& \ & \hat{\Gamma}^{(n)}_{\mu\nu} \hat{\Gamma}^{(n)}_{\nu} = 4
\hat{\Gamma}^{(n)}_{\mu} \nonumber \\
& \ & \hat{\Gamma}^{(n)}_{\mu\nu} \hat{\Gamma}^{(n)}_{\nu\mu} = 4
n ( n + 4 ) \nonumber \\
& \ & \hat{\Gamma}^{(n)}_{\mu\nu} \hat{\Gamma}^{(n)}_{\nu\lambda}
= n ( n + 4 ) \delta_{\mu\lambda} + \hat{\Gamma}^{(n)}_{\mu}
\hat{\Gamma}^{(n)}_{\lambda} - 2 \hat{\Gamma}^{(n)}_{\lambda}
\hat{\Gamma}^{(n)}_{\mu} . \label{29}
\end{eqnarray}

Unfortunately, the basic coordinate operators $\hat{x}_{\mu}$ do
not close between them but we need to enlarge the algebra to the
commutators :

\begin{eqnarray}
& \ & \hat{\Gamma}^{(n)}_{\mu\nu} = \frac{1}{2} [
 \hat{\Gamma}^{(n)}_{\mu}, \hat{\Gamma}^{(n)}_{\nu} ] \nonumber \\
& \ & [ \hat{\Gamma}^{(n)}_{\mu}, \hat{\Gamma}^{(n)}_{\nu\lambda}
] = 2 ( \delta_{\mu\nu} \hat{\Gamma}^{(n)}_{\lambda} -
\delta_{\mu\lambda} \hat{\Gamma}^{(n)}_{\nu} ) \nonumber \\
& \ & [ \hat{\Gamma}^{(n)}_{\mu\nu},
\hat{\Gamma}^{(n)}_{\lambda\rho} ] = 2 ( \delta_{\nu\lambda}
\hat{\Gamma}^{(n)}_{\mu\rho} + \delta_{\mu\rho}
\hat{\Gamma}^{(n)}_{\nu\lambda} - \delta_{\mu\lambda}
\hat{\Gamma}^{(n)}_{\nu\rho} - \delta_{\nu\rho}
\hat{\Gamma}^{(n)}_{\mu\lambda} ) . \label{210} \end{eqnarray}

These form the $SO(5,1)$ algebra.

Between all the possible representations of this algebra, we can
always choose to diagonalize a matrix $\hat{\Gamma}_{\mu}$ out of
the five matrices, for example $\hat{x}_5= \rho \hat{\Gamma}_5$,
whose eigenvalue $\Gamma_5$ runs between $ n, n - 2, ...., - n +
2, - n $.

The matrices $\hat{\Gamma}^{(n)}_{\mu\nu}$ , with $ \mu, \nu = 1,
... 4$ form an $SO(4)$ algebra, which is a subalgebra of the
$SO(5)$ algebra :

\begin{eqnarray}
& \ & [ \hat{N}_i , \hat{N}_j ] = i \epsilon_{ijk} \hat{N}_k
\nonumber \\
& \ & [ \hat{M}_i , \hat{M}_j ] = i \epsilon_{ijk} \hat{M}_k
\nonumber \\
& \ & [ \hat{N}_i , \hat{M}_j ] = 0 \label{211}
\end{eqnarray}

where
\begin{eqnarray}
& \ & \hat{N}_1 = - \frac{i}{4} ( \hat{\Gamma}_{23} -
\hat{\Gamma}_{14} ) \ \ \ \ \ \ \  \ \hat{M}_1 = - \frac{i}{4} (
\hat{\Gamma}_{23} + \hat{\Gamma}_{14} ) \nonumber \\
& \ & \hat{N}_2 = \frac{i}{4} ( \hat{\Gamma}_{13} +
\hat{\Gamma}_{24} ) \ \ \ \ \ \ \  \ \hat{M}_2 = \frac{i}{4} (
\hat{\Gamma}_{13} - \hat{\Gamma}_{24} ) \nonumber \\
& \ & \hat{N}_3 = - \frac{i}{4} ( \hat{\Gamma}_{12} -
\hat{\Gamma}_{34} ) \ \ \ \ \ \ \  \ \hat{M}_1 = - \frac{i}{4} (
\hat{\Gamma}_{12} + \hat{\Gamma}_{34} ) . \label{212}
\end{eqnarray}

$\hat{\Gamma}_{ab}$ is rewritten as:
\begin{eqnarray}
\hat{\Gamma}_{23} = 2i ( \hat{N}_1 + \hat{M}_1 ) \ \ \ \ \ \
\hat{\Gamma}_{14} = - 2i ( \hat{N}_1 - \hat{M}_1 ) \nonumber \\
\hat{\Gamma}_{13} = - 2i ( \hat{N}_2 + \hat{M}_2 ) \ \ \ \ \ \
\hat{\Gamma}_{24} = - 2i ( \hat{N}_2 - \hat{M}_2 ) \nonumber \\
\hat{\Gamma}_{12} = 2i ( \hat{N}_3 + \hat{M}_3 ) \ \ \ \ \ \
\hat{\Gamma}_{34} = - 2i ( \hat{N}_3 - \hat{M}_3 ) . \label{213}
\end{eqnarray}

The Casimir of each $SU(2)$ algebra depends on $n$ and on the
eigenvalue $\Gamma_5$:

\begin{eqnarray}
& \ & \hat{N}_i \hat{N}_i = \frac{1}{16} ( n + \Gamma_5 ) ( n + 4
+ \Gamma_5 ) \nonumber \\
& \ & \hat{M}_i \hat{M}_i = \frac{1}{16} ( n - \Gamma_5 ) ( n + 4
- \Gamma_5 ) . \label{214}
\end{eqnarray}

Therefore matrices $\hat{N}_i$ and $\hat{M}_i$ are realized by
$(\frac{n + \Gamma_5 + 2 }{2})$ and $(\frac{n - \Gamma_5 + 2
}{2})$ dimensional representations of $SU(2)$ respectively.

For $\Gamma_5 = n$ at the north pole , the Casimir of $\hat{N}_i$
and $\hat{M}_i$ are given by :

\begin{eqnarray} & \ & \hat{N}_i \hat{N}_i = \frac{ n ( n + 2 ) }{4} \nonumber \\
& \ & \hat{M}_i \hat{M}_i = 0 . \label{215}
\end{eqnarray}

A fuzzy two-sphere appears at the north pole and it is given by
the $(n+1)$-dimensional representation of $SU(2)$. By using the
$SO(5)$ symmetry we can attach a fuzzy two-sphere to every point
on the fuzzy four-sphere. We can regard this two-sphere as the (
spin ) internal two-dimensional space.

It is not difficult to introduce actions on a fuzzy four-sphere,
for example by introducing the following matrix model:

\begin{eqnarray} S & = & - \frac{1}{g^2} Tr ( \frac{1}{4} [ A_{\mu}, A_{\nu} ] [
A_{\mu}, A_{\nu} ] + \frac{2\lambda}{5 ( n + 2 ) \rho}
\epsilon^{\mu\nu\lambda\rho\sigma}  A_{\mu} A_{\nu} A_{\lambda}
A_{\rho} A_{\sigma} \nonumber \\
& + & 8 ( 1 - \lambda ) \rho^2 A_{\mu} A_{\mu} ) \label{216}
\end{eqnarray}

which contains the fuzzy four sphere as a classical solution :

\beq A_{\mu} = \hat{x}_{\mu} = \rho \hat{\Gamma}^{(n)}_{\mu} .
\label{217} \eeq

By expanding matrices around the classical solution corresponding
to the fuzzy four-sphere, we obtain an action of noncommutative
gauge theory.

\section{Hopf fibration and projectors for the fuzzy four-sphere }

Instead of searching for instanton solutions on the fuzzy
four-sphere by exploring the classical solutions of the action
(\ref{216}), as we did for the fuzzy sphere \cite{13}, we appeal
to another method which is simpler and is based on the Hopf
principal fibration from $S^7$ to $S^4$.

In two dimensions noncommutative monopoles over the fuzzy
two-sphere were derived by introducing projector operators $P_n$
which characterize the non trivial bundles over the fuzzy
two-sphere. The canonical connection associated with the projector
$P_n(x)$ has curvature given by

\beq \nabla^2 = P_n dP_n dP_n . \label{31} \eeq

The projector operator is made in terms of the Hopf fibration from
$S^3$ to $S^2$ as follows:

\begin{eqnarray}
& \ & x_1 = \frac{\rho}{2} ( z_0 \overline{z}_1 + z_1
\overline{z}_0 ) \ \ \
\ \ \ \ \  \ \ |z_0|^2 + |z_1|^2 = 1 \nonumber \\
& \ & x_2 = \frac{\rho}{2} ( z_0 \overline{z}_1 - z_1
\overline{z}_0 ) \ \ \
\ \ \ \ \  \ \ x^2_1 + x^2_2 + x^2_3 = \frac{\rho^2}{4} \nonumber \\
& \ & x_3 = \frac{\rho}{2} ( z_0 \overline{z}_0 - z_1
\overline{z}_1 ) . \label{32}
\end{eqnarray}

In this classical setting, $z_0$ and $z_1$ are two complex numbers
constrained to be $S^3$ and the Hopf fibration produces the real
coordinates $x_i$ which are instead constrained to be $S^2$.

This representation is particularly useful for the noncommutative
case since generalizing the complex coordinates to a couple of
oscillators:

\beq [ z_i , \overline{z}_j ] = \delta_{ij} \ \ \ \ \  \ \  \ [
z_i , z_j ] = 0 \label{33}\eeq

produces for the real coordinates $x_i$, promoted to operators,
the more complex algebra of the fuzzy two-sphere:

\beq [ \hat{x}_i, \hat{x}_j ] = i \rho \epsilon_{ijk} \hat{x}_k .
\label{34} \eeq

Introducing the number operator

\beq \hat{N} = \overline{z}_0 z_0 + \overline{z}_1 z_1 \label{35}
\eeq

the parameter $\rho$ is given in terms of $n$, the eigenvalue of
$\hat{N}$, and $R$ the radius of the sphere:

\begin{eqnarray}
& \ & \sum_i {( \hat{x}^i )}^2 = R^2 \nonumber \\
& \ & \rho = \frac{2R}{\sqrt{ n ( n + 2 )}} . \label{36}
\end{eqnarray}

Here is the idea of the present letter. Since the Hopf fibration
$S^3 \rightarrow S^2$ is generalizable to the case $S^7
\rightarrow S^4$, is it possible to represent the complicated
algebra of the fuzzy four-sphere in terms of four independent
oscillators ? The answer is yes, and it is also possible to
generalize the construction that led us to compute the projectors
$P_n$ for the fuzzy two-sphere to the case of fuzzy four-sphere.

The Hopf fibration $S^7 \rightarrow S^4$ is made by four complex
coordinates $z_i$ constrained to be $S^7$, that are mixed together
to give five real coordinates $x_i$ constrained to be $S^4$:

\begin{eqnarray}
& \ & x_1 = \rho ( a_1 + \overline{a}_1 ) \ \ \ \ \ \ \ \ x_2 = i
\rho ( a_1 - \overline{a}_1 ) \nonumber \\
& \ & x_3 = \rho ( a_2 + \overline{a}_2 ) \ \ \ \ \ \ \ \ x_4 = i
\rho ( a_2 - \overline{a}_2 ) \nonumber \\
& \ & x_5 = \rho ( z_0 \overline{z}_0 + z_1 \overline{z}_1 - z_2
\overline{z}_2 - z_3 \overline{z}_3 ) \nonumber \\
& \ & a_1 = z_0 \overline{z}_2 + z_3 \overline{z}_1 \nonumber \\
& \ & a_2 = z_0 \overline{z}_3 - z_2 \overline{z}_1 \nonumber \\
& \ & \sum_i x^2_i = \rho^2 \ \ \ \ \ \ \  \ \sum_i |z_i|^2 = 1 .
\label{37}
\end{eqnarray}

By promoting the complex coordinates $z_i$ to four oscillators as
follows :

\beq [ z_i , \overline{z}_j ] =  \delta_{ij} \ \ \  \ \ \ [ z_i,
z_j ] = 0 \label{38} \eeq

the corresponding algebra for $\hat{x}_i$ is the fuzzy four-sphere
algebra. For example the $SU(2) \times SU(2)$ subalgebra made by
$\hat{N}_i$ and $\hat{M}_i$ can be easily represented in terms of
$z_i$ as:

\begin{eqnarray}
& \ & \hat{N}_3 = \half ( z_3 \overline{z}_3 - z_2 \overline{z}_2
) \ \ \ \ \ \ \ \ \ \ \ \hat{M}_3 = \half ( z_0 \overline{z}_0 -
z_1
\overline{z}_1 ) \nonumber \\
& \ & \hat{N}_{+} = \hat{N}_1 + i \hat{N}_2 = z_2 \overline{z}_3 \
\ \ \ \  \ \ \ \ \ \hat{M}_{+} = \hat{M}_1 + i \hat{M}_2 = z_1
\overline{z}_0 \nonumber \\
& \ & \hat{N}_{-} = \hat{N}_1 - i \hat{N}_2 = z_3 \overline{z}_2 \
\ \ \ \  \ \ \ \ \ \hat{M}_{-} = \hat{M}_1 - i \hat{M}_2 = z_0
\overline{z}_1 . \label{39}
\end{eqnarray}

The parameter $n$ is simply the eigenvalue of the number operator
$\hat{N}$ :

\beq \hat{N} = \overline{z}_0 z_0 + \overline{z}_1 z_1 +
\overline{z}_2 z_2 + \overline{z}_3 z_3 \ \ \ \ \ \ \hat{N}
\rightarrow n  . \label{310} \eeq

In terms of $\hat{N}$, the Casimir for $\hat{x}^2_i$ is :

\beq \sum_i \hat{x}^2_i = \rho^2 \hat{N} ( \hat{N} + 4 ) = R^2 .
\label{311} \eeq

To construct the $k$-instanton projectors $P_k(x)$ for the fuzzy
four-sphere, let us consider the following vectors:

\beq | \psi_k > = N_k \left( \begin{array}{c} (z_0)^k \\
........ \\
\sqrt{ \frac{ k ! }{ i_1 ! i_2 ! i_3 ! ( k - i_1 - i_2 - i_3 ) !
}}
z_0^{k - i_1 - i_2 - i_3} z_1^{i_1} z_2^{i_2} z_3^{i_3} \\
........  \\
(z_1)^k \end{array} \right) \label{312} \eeq

where

\beq 0 \leq i_1 \leq k \ \ \ \ \ \ 0 \leq i_2 \leq k - i_1 \ \ \ \
\ \ 0 \leq i_3 \leq k - i_1 - i_2  . \label{313} \eeq

Fixing $i_1$ and $i_2$, the index $i_3$ takes $ k - i_1 - i_2 +1 $
values. Fixing $i_1$, the indices $i_2$ and $i_3$ take $\frac{( k
- i_1 + 1 )( k - i_1 + 2 )}{2}$ values.

In total the number of entries of the vector $|\psi_k >$ is given
by summing over $i_1$ as follows:

\beq N_k = \frac{ ( k + 1 ) ( k + 2 ) ( k + 3 ) }{6} \label{314}
\eeq

which is equal to the number $N$ of the size of the matrix
$\hat{\Gamma}^{(k)}_{\mu}$.

The normalization condition for these vectors fixes the function
$\hat{N}_k$ to be dependent only on the number operator $\hat{N}$:

\begin{eqnarray}
& \ & < \psi_k | \psi_k > = 1 \nonumber \\
& \ & N_k = N_k ( \hat{N} ) = \frac{1}{\sqrt{\prod_{i=0}^{k-1} (
\hat{N} - i + k ) }} = \frac{1}{\sqrt{\prod_{i=0}^{k-1} ( n - i +
k ) }}  . \label{315}
\end{eqnarray}

The corresponding $k$-instanton connection $1$-form can be
computed in terms of the vector $|\psi_k>$:

\beq A_k^{\nabla} = < \psi_k | d |\psi_k > . \label{316} \eeq

The projector for the $k$-instanton on the fuzzy four sphere is
defined to be:

\beq P_k = | \psi_k >< \psi_k | \ \ \ \ \ \ \ \ P^2_k = P_k \ \ \
\ P^{\dagger}_k = P_k . \label{317} \eeq

In the product of the two ket and bra vectors, the sixteen
combinations of oscillators can be written in terms of the algebra
of the fuzzy sphere and the number operator, which is equal to its
eigenvalue $n$ on a definite representation. Therefore $P_k$ is a
matrix having as entries the basic operator algebra of the theory.

The trace of the projector $P_k$ is always positive definite since

\begin{eqnarray} & \ &
Tr P_k = \frac{ ( n + k + 1 )( n + k + 2 ) ( n + k + 3 ) }{( n + 1
) ( n + 2 ) ( n + 3 )} Tr 1 = \nonumber \\
& \ & = \frac{ ( n + k + 1 )( n + k + 2 ) ( n + k + 3 ) }{6} <
\frac{ ( n + 1 )( n + 2 ) ( n + 3 ) }{6} \frac{ ( k + 1 )( k + 2
)( k + 3 ) }{6} = Tr 1_P \nonumber \\
& \ & \label{318} \end{eqnarray} where $1_P$ is the identity
projector.

To construct the $k$-anti-instanton solution it is enough to take
the adjoint of the vector $|\psi_k>$. Consider the $\frac{( k +
1)( k + 2 )( k + 3 )}{6}$-dimensional vectors:

\beq | \psi_{-k} > = N_k \left( \begin{array}{c} (\overline{z}_0)^k \\
........ \\
\sqrt{ \frac{ k ! }{ i_1 ! i_2 ! i_3 ! ( k - i_1 - i_2 - i_3 ) !
}} \overline{z}_0^{k - i_1 - i_2 - i_3} \overline{z}_1^{i_1}
\overline{z}_2^{i_2}
\overline{z}_3^{i_3} \\
........  \\
(\overline{z}_1)^k \end{array} \right) . \label{319} \eeq

The normalization condition for these vectors fixes the function
$N_k$ to be dependent only on the number operator $\hat{N}$:

\begin{eqnarray}
& \ & < \psi_{-k} | \psi_{-k} > = 1 \nonumber \\
& \ & N_k = N_k ( \hat{N} ) = \frac{1}{\sqrt{\prod_{i=0}^{k-1} (
\hat{N} + i + 4 - k ) }} = \frac{1}{\sqrt{\prod_{i=0}^{k-1} ( n +
i + 4 - k ) }} . \label{320}
\end{eqnarray}

The corresponding projector for anti-instantons is

\beq P_{-k} = |\psi_{-k} ><\psi_{-k} | \ \ \ \ \ \\  k < n + 4 .
\label{321} \eeq

The trace of this projector

\begin{eqnarray} & \ & Tr P_{-k} =
\frac{ ( n - k + 1 )( n - k + 2 ) ( n - k + 3 ) }{( n + 1
) ( n + 2 ) ( n + 3 )} Tr 1 = \nonumber \\
& \ & = \frac{ ( n - k + 1 )( n - k + 2 ) ( n - k + 3 ) }{6}
\label{322}
\end{eqnarray}

is positive definite if and only if the following bound is
respected

\beq  k < n + 1 . \label{323} \eeq

For the special cases $ k = n + 1, k = n + 2, k = n + 3 $ $P_{-k}$
is simply the null projector.

The projector above, defining the $k$-projective moduli of the non
commutative gauge theory over the fuzzy four-sphere can be used to
compute the following Chern class

\beq c_k = Tr ( \gamma_5 P_k d P_k d P_k d P_k d P_k ) \label{324}
\eeq

as we did in \cite{12} for the fuzzy two-sphere.

\section{Conclusions}

In this paper we have made use of an alternative description to
the instanton connections and gauge fields in terms of globally
defined projectors, a method which is suitable for generalization
like noncommutative geometry.

In particular we have shown how to derive exact expressions for
the noncommutative $k$-instantons on the fuzzy four-sphere. The
projectors defined here are finite-dimensional matrices as in the
two dimensional case, and can be used to compute the corresponding
topological charge ( Chern class for the vector bundle ) in four
dimensions. We believe that this intrinsic description of
connections and gauge fields in global terms will help in
extending the methods of noncommutative geometry to the case of
the fuzzy four-sphere.

A possible generalization of this paper would be to study the flat
limit of the fuzzy four-sphere, which should define the
noncommutative $k$-instantons on the noncommutative four plane.
This could represent an alternative method with respect to the
papers
\cite{14}-\cite{15}-\cite{16}-\cite{17}-\cite{18}-\cite{19}-\cite{20},
based on the Hopf fibration $S^7 \rightarrow S^4$ and the flat
limit.

\end{document}